\documentstyle[aps,prl,preprint]{revtex}

\let\ssection=\section
\renewcommand{\section}{\setcounter{equation}{0}\ssection}

\begin{document}
\draft

\preprint{IMPERIAL /TP/93-94/12}

\title{Sharpened Information-Theoretic Uncertainty Relations and the
Histories Approach to Quantum Mechanics}
\author{Bernhard Meister}
\address{Blackett Laboratory, Imperial College, London SW7 2BZ,
United Kingdom}
\date{\today}

\maketitle

\begin{abstract}
In this paper alternative formulations of the conventional
  uncertainty relation
are studied in the  context of decoherent histories. The
results are given in terms of Shannon information.
A variety of methods are developed to evaluate the
upper bound for the probability of  two or more projection histories.
The methods employed give  improved  limits for  the maximal achievable
probability and an improved lower bound  for the Shannon information.
The results are then applied to a  number  of physically
 relevant situations.
\end{abstract}
\pacs{PACS Numbers: 03.65.Bz, 02.50.Cw, 05.20.-y}

\section{Introduction}


In this paper, we will explore some mathematical aspects of
 the formulation of quantum mechanics based on the idea of a
quantum-mechanical history,
  defined to be a
time-ordered sequence
of projection operators acting on an initial state.
The projection operators at each moment of time satisfy
\begin{eqnarray}
P_{\alpha}P_{\beta}=\delta_{\alpha\beta}\,\, P_{\alpha}\, ,
\end{eqnarray}
and
\begin{eqnarray}
\sum_{\alpha}P_{\alpha}=1\, .
\end{eqnarray}

The probability of a n-projection history with  an initial state represented by
a
density matrix $\rho$ is given by
\begin{eqnarray}
p(\alpha_1,\cdots,
\alpha_n)=Tr[P^{n}_{\alpha_{n}}(t_{n})\cdots P^{1}_{\alpha_{1}}(t_{1})\rho
P^{1}_{\alpha_{1}}(t_{1})\cdots P^{n}_{\alpha_{n}}(t_{n})]\label{eq:p}
\end{eqnarray}
with the time evolution operator $e^{-iHt/\hbar}$ allowing us to
write for the m'th-projection
\begin{eqnarray}
P^{m}_{\alpha_m}(t_{m})=e^{i H (t_{m}-t_0)/\hbar}P^{m}_{\alpha_m}(t_0)e^{-i H
(t_{m}-t_0)/\hbar}\, .
\end{eqnarray}
A general form of this  probability functional  plays a central role in the
decoherent histories approach.
For more details about the decoherent histories approach see \cite{gri} and
references therein.
 An axiomatic foundation of the
decoherent histories approach is given in \cite{ha},\cite{is}.
In standard  quantum mechanics, the  uncertainty principle
takes the form
\begin{eqnarray}
\Delta p \Delta q \geq \frac{\hbar}{2}\, .
\end{eqnarray}
It relies very much on the notion of the state at a fixed moment of time.
In this paper we are concerned with the question of how the
uncertainty principle arises in a formulation based on the
expression for the probability of histories, (\ref{eq:p}).
This question has been addressed by Halliwell \cite{jjh1}, who argued that the
uncertainty principle may arise as a lower bound on the
Shannon information
  of the probability (\ref{eq:p}). He derived  approximate although explicit
 expressions for the lower bound in certain situations.

The  aim of the present paper is to obtain more accurate and more general
 versions of the known information-theoretic inequalities by introducing more
sophisticated
mathematical
techniques which  are likely to be of use in a wide
variety of related problems.
The other main aim is to understand how information can be used to
measure uncertainty in standard and  generalized versions of
quantum mechanics \cite{is}. It  means that the role of
information theory has
 to be reconsidered in the common
understanding of  quantum mechanics.

In Section II we  introduce Shannon information and briefly describe
some of its properties.
The next few sections are taken up with the development
of  methods for calculating the maximum probabilities of
quantum-mechanical histories.
These results are then used in section VII to obtain upper and lower
bounds for the maximum probabilities in a  variety of situations.
Section VIII is used to calculate the upper bounds for the probabilities of
N-projection histories.
In section IX the results of the previous sections
are applied to gain a lower bound for Shannon information based on
$I\geq -\log (p_{max})$.

There are two more points of interest:
 This paper is motivated  by  the decoherent histories
approach, but all
the calculations are totally independent of it and can be carried
out in standard quantum mechanics.

Secondly we are mainly concerned with physically realizable
projections and quantum-mechanically (experimentally) interesting
situations. We will later  make a statement about the physical relevance
of approximate (Gaussian) projections.
In this regime  $\sigma_x \sigma_y M/2  \hbar t$, $\sigma_x \sigma_k/ 2  \hbar
$
 or similar constants are small. The constants used are  the mass $M$, the time
separation between the projections $T$ and the size of the exact
position and momentum projections $\sigma_x,  \sigma_y$ and $\sigma_k$.

\section{Information Theory}\label{2}

We give a short introduction to certain aspects of
Shannon information theory \cite{sh},\cite{co1}
which are useful in finding alternative formulations of the
standard quantum-mechanical uncertainty principle.  The information theoretic
approach has
been used extensively  in  \cite{ar} and  \cite{jjh1}.

In the discrete case we have a probability  function $p(n)$
defined on the set $n=1,...,N$, satisfying $0\leq p(n) \leq 1$ and
\begin{eqnarray}
\sum_{n=0}^{N} p(n)=1\, .
\end{eqnarray}
In the continuous case we have
\begin{eqnarray}
\int dx\, p_c(x)=1 \, .
\end{eqnarray}
 The corresponding information  in the discrete case is defined as
\begin{eqnarray}
I_{d}=-\sum_{n}\, p(n)\log p(n)\nonumber
\end{eqnarray}
and  the continuous case it is
 \begin{eqnarray}
I_{c}=-\int  dx \,p_c(x)\log p_c(x)\nonumber \, .
\end{eqnarray}
The relationship between discrete and continuous information can
be understood best by  studying an example.
In particular for  a Gaussian
position projection the  two kinds of information are related by
\begin{eqnarray}
I_{d}(\bar{X})&\equiv& -\sum_{n} p(n)\log p(n)\nonumber\\
&\geq& -\int dx \langle x|\rho|x\rangle \log \langle x|\rho|x\rangle
-\log\sigma_{\alpha}\nonumber\\
&\equiv& I_{c}(X)-\log (\sigma_{\alpha})\nonumber
\end{eqnarray}
with
 \begin{eqnarray}
 p(n)=Tr[P^{x}_{n}\rho ] \, .
 \end{eqnarray}
and $\sigma_{\alpha}$ the width of the Gaussian sampling function.
This case is discussed and the notation explained in detail in \cite{jjh1}.

In the discrete case an upper and lower bound for $I_d$ can easily be given:
\begin{eqnarray}
0 \leq I_d \leq \log N \, .
\end{eqnarray}
The upper bound is reached for $p(n)$ equal to $1/N$ for all $n$.

\section{Two-Time Histories}
For most of this paper we will concentrate on the case of histories
characterized by position and/or momentum projections at two moments
of time. The probability for a two-time history can be written as
\begin{eqnarray}
p( \alpha,  \beta, T )
= Tr(P_{\alpha} e^{iHT} P_{ \beta} e^{-iHT} P_{\alpha}\rho)\, .
\end{eqnarray}

Before we seek to maximize the value of the
probability
over all alternatives $\alpha$,  $\beta$, and over all initial
states $\rho$, we will demonstrate with the help   of Weyl-Heisenberg coherent
states that we can choose $\alpha$ and $\beta$ arbitrarily without
changing the maximum probability.
This means   the
 maximum probability of any two-projection history is
translation invariant, depending only
 on the size of the projections, but not on their  positions
in space.
 This will give us a lower bound for the Shannon information.
 This result has been derived previously by Halliwell \cite{jjh1}
 using  the Wigner transform. Our method is not only
simpler, but can also be generalized to a large variety of physical situations.

In the case of $\rho=\vert \psi \rangle\langle\psi \vert$ there is a set of
wavefunctions $\psi_{max}$ which maximize the probability.
We can restrict $\rho$ to be a  pure state without changing the upper
bound for the probability.
If the wave function $\psi_{max}$, achieving maximal probability for
one
 particular configuration, is known, then
$U({\mathaccent 22 x},{\mathaccent 22 k})\psi_{max}$ is the
corresponding wave-function in the new translated  situation.

\subsection{Two-Position Histories}
We study the probability of a history consisting of position
samplings at two moments of time.
In this case $P_{\alpha_1}$ and $P_{\alpha_2}$ are two position projections
and  the probability-functional  has the form
\begin{eqnarray}
p(\alpha_1,\alpha_2, T ) = Tr \left[ P^x_{\alpha_1} e^{-iHT}
P^x_{\alpha_2} \rho P^x_{\alpha_2}
e^{iHT} \right]
\end{eqnarray}
where the projection  has the general form
\begin{eqnarray}
P^x_{\alpha}=\int  dx \Upsilon (x-{\mathaccent 22 x}_{\alpha})
|x\rangle \langle x|
\end{eqnarray}
with the sampling function satisfying
\begin{eqnarray}
\int dx \Upsilon (x-{\mathaccent 22 x}_{\alpha})&=&\sigma_x\\
\sum_{\alpha} \Upsilon (x-{\mathaccent 22 x}_{\alpha})&=&1\,
\end{eqnarray}
and
\begin{eqnarray}
 {\mathaccent 22 x}_{\alpha}=\alpha \sigma_{x}\,.
\end{eqnarray}
For exact projections the sampling function $\Upsilon
(x-{\mathaccent 22 x}_{\alpha})$
 is
\begin{eqnarray}
\Theta
\Big( \frac{x-{\mathaccent 22 x}_{\alpha}+\frac{1}{2}\sigma_x}{\sigma_x}\Big)
\Theta\Big(\frac{-x+{\mathaccent 22
x}_{\alpha}+\frac{1}{2}\sigma_x}{\sigma_x}\Big)
\end{eqnarray}
where $\sigma_x$ is the width,  ${\mathaccent 22 x}_{\alpha}$ is the center of
the exact projection and $\alpha$ is an integer.
Now we will  show that $p(\alpha_1 ,\alpha_2 , T )$ is equal to
$Tr(U({\mathaccent 22 k},{\mathaccent 22 x})
\Omega_{x}U({\mathaccent 22 k},{\mathaccent 22 x})\rho )$
where
\begin{eqnarray}
\Omega_{x} = \frac{1 }{ 2 \pi \hbar} \int dx dy dy' \Upsilon (x) \Upsilon ( y)
\Upsilon (y')
 e^{i\frac{m}{2\hbar T} [(x-y)^2-  (x-y')^2]}
|y\rangle \langle y'|\, .
\end{eqnarray}
This can be done with the help of Weyl-Heisenberg coherent-states.
We define:
\begin{eqnarray}
U(p,q)&=&e^{\frac{i}{\hbar}(p{\hat Q}-q{\hat P})}\\
U(p,q)|x\rangle&=&|x+q\rangle e^{\frac{i}{\hbar}p(x+q)}\\
U(p,q)|k\rangle&=&|k+p\rangle e^{\frac{i}{\hbar}q(k-p)}
\end{eqnarray}
  The  commutation relation is
\begin{eqnarray}
e^{{\hat Q}+{\hat P}}=e^{{\hat Q}}e^{{\hat P}}e^{-[{\hat Q},{\hat P}]/2}
\end{eqnarray}
and the free particle-evolution is given by
\begin{eqnarray}
\langle x,t|y,0 \rangle = \sqrt{\frac{m}{2\pi \hbar i t}} \exp
\Big(\frac{im}{2\hbar t} (x-y)^2\Big)
\end{eqnarray}
which we assume to be a good {\it approximation} in general and exactly true
for  the cases we analyse in the next two sections\footnote{ The proof for
cases analysed  in later sections
follows trivially from the one given above.}.
In this situation our history is
$P_{{\mathaccent 22 x}}^{x} e^{i H t} P_{{\mathaccent 22 y}}^{x}$
and  we set
\begin{eqnarray}
{\mathaccent 22 k}
 =
\frac{({\mathaccent 22 x}-{\mathaccent 22 y})m}{t}\, .
\end{eqnarray}
By writing the equation for the probability
\begin{eqnarray}
Tr (U({\mathaccent 22 k},{\mathaccent 22 x})\Omega_x U^{\dagger}({\mathaccent
22 k},{\mathaccent 22
x})\rho)
\end{eqnarray}
 explicitly and doing a straightforward substitution, the translational
invariance of the maximum probability  is derived.

\subsection{Momentum-Position History}
For a position-momentum history the probability $p(\sigma_x,\sigma_k,T)$ is
given by the integral
\begin{eqnarray}
 \frac{1 }{ 2 \pi \hbar} \int dk \int dy \int dx \Upsilon (x -{\mathaccent
22 x}) \Upsilon ( y-{\mathaccent 22 y}) \Gamma (k-{\mathaccent 22
k})\nonumber\\
 \langle x|e^{iHT}|k\rangle \langle k|e^{-iHT}|y\rangle
  \psi(x)\psi^*(y)\nonumber
\end{eqnarray}
where $\Gamma (k-{\mathaccent 22 k})$ is the sampling function for
the momentum.
In the short time limit
the terms containing $H$ cancel out.
The goal  again is to reformulate the equation (9) containing
$P_{\alpha}$ and $P_{\beta}$
into the form
$U({\mathaccent 22 k},{\mathaccent 22 x})
\Omega_{k}U({\mathaccent 22 k},{\mathaccent 22 x})$
with
\begin{eqnarray}
\Omega_{k} = \frac{1 }{ 2 \pi \hbar} \int dx \int dy \int dk \Upsilon (x)
\Upsilon ( y) \Gamma (k)
 e^{i k(x-y)/\hbar }
|x\rangle \langle y|
\end{eqnarray}
using  $\langle x|k\rangle =e^{ikx/\hbar}$.
Applying $U$-operators  explicitly and using substitution gives the
expected result
\begin{eqnarray}
Tr[P^{x}_{\alpha} e^{-iHT}P^{k}_{\beta } e^{-iHT} P^{x}_{\alpha}\rho ]=Tr[
U({\mathaccent 22 k},{\mathaccent 22 x})\Omega_k  U^{\dagger}({\mathaccent 22
k},{\mathaccent 22 x})\rho]
 \,\, .
\end{eqnarray}

 \section{Calculation for 2 Projection Histories in the Free-Particle Case}

Now we do the explicit calculation for the free-particle case and
exact projections.
The probability is

\begin{eqnarray}
p(\sigma_x,\sigma_y,T)&=&\int_{-\frac{\sigma_x}{2}}^{\frac{\sigma_x}{2}}dx\Big\vert
\int_{0}^{\sigma_y}dy\,\, K(x,T;y,0)\psi (y,0)\Big\vert^2
\nonumber\\
&=&\frac{A'}{\pi }\int_{-\frac{1}{2}}^{\frac{1}{2}} dx\Big\vert
\int_{0}^{\sigma_y}dy \,\, e^{iA'(x-y)^2}\psi (y,0)\Big\vert^2
\nonumber
\\
&=&\int_{0}^{\sigma_y}dy
\int_{0}^{\sigma_y}  dy' \,\,
e^{iA'y^2}e^{-iA'y'^2}\frac{\sin(A'(y-y'))}{\pi(y-y')}\psi( y,0)\psi^*(
y',0)\nonumber
\end{eqnarray}
where $A'=\sigma_x  m/2  \hbar T$ and the free
particle propagator is
\begin{eqnarray}
K(x,T;y,0)=\sqrt{\frac{ m}{2 i \hbar t \pi}}\exp \Big(\frac{i m}{2 \hbar T }
(x-y)^2\Big)\, .
\end{eqnarray}
The change of notation from $p(\alpha_1,\alpha_2, T )$ to
$p(\sigma_x,\sigma_y,T)$ for the probability  is beneficial, because we
concentrate in
the next sections
 on calculating an upper bound for the probability of a two-position
history with arbitrary $\alpha_1$ and $\alpha_2$, but with fixed
$\sigma_x,\sigma_y,m$ and $T$.

To simplify the problem  a  redefinition of $\psi$ to
$\psi(y)e^{iA'y^2}$ is useful.
Next we  approximate $\psi$ by  piecewise
constant functions. This can be justified  using  the  basic ideas of
multiresolution analysis \cite{me},\cite{da}. We start with a sequence of
successive
approximation spaces $V_{N}$.
These  subspaces of $L^2(\Re)$ satisfy the following criteria:

\begin{eqnarray}
\cdots V_{-2}\subset V_{-1}\subset V_{0}\subset V_{1}\subset V_{2}\cdots
\end{eqnarray}
with
\begin{eqnarray}
\overline{\bigcup_{N\in {\it Z}} V_{N}} =L^{2}(\Re)
\end{eqnarray}
and
\begin{eqnarray}
\bigcap_{N\in {\it Z}} V_{N} ={0}\, .
\end{eqnarray}
We also impose the additional scaling requirement
that $\psi\in V_{N}$
implies, and is implied by,
$\psi(2^N \cdot)\in V_0$
and the invariance under integer translations
so that $\psi\in V_{0}$
implies, and is implied by,
$\psi(\cdot-k)\in V_0$ for all $k\in Z$.

The orthogonal projection $P_{N}$ onto $V_N$ allows us to write
\begin{eqnarray}
\lim_{N\rightarrow\infty}P_{N}\psi=\psi
\end{eqnarray}
for all $\psi\in L^{2}(\Re)$.
The simplest example for this ladder of spaces is given by
\begin{eqnarray}
V_{N}
=\{\psi\in L^2(\Re)\, ;
 \psi \rm{\,  piecewise\,\,
constant \,\, on\,\,
the\,\, half\,\,  open\,\,  interval\, }
[2^{-N} n,2^{-N}(n+1)[
,\, n\in Z\}\nonumber
\end{eqnarray}
The orthonormal basis of $V_{N}$ is
\begin{eqnarray}
\{\psi_{n,N}; n\in Z \}
\end{eqnarray}
where  $\psi_{n,N}$ is defined as
\[
\psi_{n,N}(x)=\left\{\begin{array}{ll}
1&\mbox{$n2^{-N}\leq x\leq(n+1)2^{-N}$}\\
0&\mbox{otherwise .}
\end{array}
\right.
\]
If $|| \psi||=1$ then
\begin{eqnarray}
\lim_{N\rightarrow \infty}\sum_{n\in Z}|a_{n,N}|^2 2^{-N}=1\, .
\end{eqnarray}
In our case we are interested only in the values of the
wave-function in the interval $[0,\sigma_y]$.
This allows us to restrict the range of $n$ from
 the integers  to the  natural numbers between $0$ and $2^{N}$, if
we redefine $\psi_{n,N}$ to be
\[
\psi_{n,N}(x)=\left\{\begin{array}{ll}
 1&\mbox{$n\,\sigma_y\,2^{-N}\leq
 x\leq(n+1)\,\sigma_y\,2^{-N}$}\\
0&\mbox{otherwise\, .}
\end{array}
\right.
\]
The integral can now be rewritten as
\begin{eqnarray}
p(\sigma_x,\sigma_y,T)
=\lim_{N\rightarrow \infty}
\sum _{n=0}^{2^{N}}\sum_{m=0}^{2^{N}}
\frac{1}{\pi}\int^{\sigma_y}_{0}dy
\int_{0}^{\sigma_y} dy'\nonumber\\
\frac{\sin(A'(y-y'))}{\pi(y-y')} \,\, a_{N,n}\,\, \psi_{N,n}(y)\,\,
a_{N,m}^*\,\, \psi_{N,m}^*(y')
\nonumber\\
=\lim_{N\rightarrow \infty}
\sum _{n=0}^{2^{N}}\sum_{m=0}^{2^{N}}
\frac{1}{\pi}\int_{n 2^{-N} \sigma_y}^{(n+1) 2^{-N} \sigma_y}dy
\int_{m 2^{-N} \sigma_y}^{(m+1) 2^{-N} \sigma_y} dy'\nonumber\\
\frac{\sin(A'(y-y'))}{\pi(y-y')} \,\, a_{N,n}\,\,
a_{N,m}^*\,
\end{eqnarray}
with
\begin{eqnarray}
\psi(y,0)=\lim_{N\rightarrow
\infty}\sum_{n=0}^{2^{N}}a_{n,N}\,\psi_{n,N}(x)\label{eq:bor}
\end{eqnarray}
and
\begin{eqnarray}
a_{n,N}=\int_{n 2^{-N} \sigma_y}^{(n+1) 2^{-N} \sigma_y}dy\,\, \psi
(y,0)\, .
\end{eqnarray}
The convergence of (\ref{eq:bor}) is given in the $L^2$-norm, for more details
see
Meyer \cite{me}.

Now we  can evaluate the integral  on each interval with
$a_na_m^*=2^{N}\sigma_y$:
\begin{eqnarray}
c_{\delta}&=&\frac{A2^{-N}}{
\pi}\int_{-\frac{1}{2}}^{\frac{1}{2}}dx\int_{0}^{1}dy'
\int^{\delta+1}_{\delta }dy\,\,  e^{i 2 A(y-y')x 2^{-N}}\nonumber\\
&=&\frac{1}{\pi}\int^{1}_{0}dy'\int^{\delta
+1}_{\delta}dy\,\,\frac{\sin(A(y-y')2^{-N})}{y-y'}\nonumber\\
&=&\frac{1}{\pi}\int^{\delta +1}_{\delta }dy \,\, Si(Ay2^{-N})-
Si(A(y-1)2^{-N})\nonumber\\
&=&\frac{1}{\pi}\Big\{2^N\frac{\cos(A(\delta-1)2^{-N})}{A}-2^{N+1}\frac{\cos(A\delta2^{-N})}{A}+2^N\frac{\cos(A(\delta +1)2^{-N})}{A}\nonumber\\
&&+(\delta-1)Si\Big(A(\delta-1)2^{-N}\Big)
-2\delta Si\Big(A\delta 2^{-N}\Big)+(\delta+1)
Si\Big(A(\delta+1)2^{-N}\Big)\Big\}\nonumber
\end{eqnarray}
with $\delta =n-m$, $c_{nm}=c_{n-m}$ and $A:=A'\sigma_y$.
This integral is solved most easily by substituting  new variables for
$y+\delta y'$ and $y-\delta y'$.
We also define
\begin{eqnarray}
Si(k):=\int_{0}^{k}dx\,\frac{\sin(x)}{x}
\end{eqnarray}
which can be approximated by
\begin{eqnarray}
\sum_{n=0}^N (-1)^{n}\frac{k^{2n+1}}{(2n+1)!\,(2n+1)}\, .
\end{eqnarray}

This means our integral-operator has been transformed into  matrix form.
Instead of $\int dy\int dx \psi(x)f(x-y)\psi^*(y)=\lambda $, we now
work with its discretized version: $\sum_{n,m}b_nC_{nm}b^*_m =$ $\lambda$ with
$C_{nm}=c_{n-m}$ with
$
\sum_{n=0}^{2^N}|b_n|^2=1$.
The bilinear
form is maximal if
$b_m$ is an eigenvector of $C_{nm}$.

This allows us to use two simple approximations for the eigenvalues
of the matrix in the case when $A<\frac{\pi}{2}$, because the matrix
coefficients are all positive.


 We know that for any matrix $C_{nm}$ all
eigenvalues are smaller than
\begin{eqnarray}
 \max_{j} \sum_{i=0}^{2^N} |C_{ij}|
\end{eqnarray}
In our case this gives as an upper bound for the eigenvalue for
$N\rightarrow \infty$ of $\frac{2}{\pi} Si(\frac{A}{2})$.
This is gained by letting  $N\rightarrow \infty$ in the sum
\begin{eqnarray}
\sum_{n=-2^{N-1}}^{2^{N-1}} c_{n}\, .
\end{eqnarray}
The upper bound for the probability is
\begin{eqnarray}
p(\sigma_x,\sigma_y,T)\leq \frac{2}{\pi}Si\Big(\frac{A}{2}\Big)\, .
\end{eqnarray}

\subsection{T\"oplitz-Method}
The next step is to study  the eigenvalues by exploiting the fact that $c_{nm}$
is a T\"oplitz matrix \cite{sc}, \cite{ka}, \cite{gr}.
The characteristic property of a T\"oplitz matrix $c_{nm}$ is the
equality of the coefficients on the diagonals.

\begin{eqnarray}
 K_{}=\left(\begin{array}{ccccc}
  c_0 & c_1 & c_2 & \cdots & c_{n-1}\\
 c_{-1}& c_0 & c_1&\cdots  &c_{n-2}\\
c_{-2} & c_{-1} & c_{0} & \cdots & c_{n-3}\\
\cdot  & \cdot & \cdot & \cdots & \cdot\\
c_{-n+1} &c_{-n+2}  &c_{-n+3}
& \cdots & c_0\end{array}\right)
\end{eqnarray}
It is also normally assumed $c_n=c^{*}_{-n}$ , but in our case
$c_n=c^{*}_{-n}=c^{*}_{n}$.
This allows us to view the $c_{n}$ as Fourier-coefficients of a real
function:
\begin{eqnarray}
c_{n}=\frac{1}{2\pi}\int^{\pi}_{-\pi} d\Theta \,\, f(\Theta)
e^{-in\Theta}\,\,\,\, n=0,\pm1,\pm2, ...
\end{eqnarray}

\begin{eqnarray}
f_{N}(\Theta)=\sum_{n=-2^{N}}^{2^{N}
}c_{n,N}e^{in\Theta}
\end{eqnarray}
and
\begin{eqnarray}
f_{N}(\Theta)=c_0+\sum_{n=1}^{2^{N}
}2c_{n,N}\cos(n\Theta)
\end{eqnarray}
or, for fixed $N$,
\begin{eqnarray}
f_{N}(\Theta)=\lim_{L\rightarrow \infty}\sum_{n=-L}^{L
}c_{n,N}e^{in\Theta}\, .
\end{eqnarray}
\vspace{1cm}

For any fixed $N$ we assume
\begin{eqnarray}
m_N\leq f_{N}(\Theta) \leq M_N
\end{eqnarray}
for all $ \Theta\in [- \pi, \pi]$.
Then we know that the eigenvalues $\lambda$ of the matrix $C_{nm}$  for a
fixed $N$ satisfy

\begin{eqnarray}
m_N\leq \lambda_1 \leq \cdots\leq\lambda_L\leq M_N\, .
\end{eqnarray}

Now to complete the process we study $A>\frac{\pi}{2}$. In this case
some of the elements of the T\"oplitz matrix are negative and the
function $f_{N}$ does not automatically reach its maximum  at $\Theta=0$.
One has to   study  $f_{N}'(\Theta)$. In the case of $N$ very large
we approximate $c_{n}$ by $ \sin(\frac{nA}{2^N})/n\pi$ and  get
\begin{eqnarray}
f'(\Theta)=\sum_{n=1}^{2^N} -n 2c_{n}\sin(n\Theta)\approx \sum_{n=1}^{2^N}
-2\sin\Big(\frac{nA}{2^N}\Big)\frac{\sin(n\Theta
)}{\pi}\, .
\end{eqnarray}

As mentioned before, this case is of secondary importance and will
not be studied in any  detail.

\subsection{Perron-Frobenius Theorem}
In this sub-section  the Perron-Frobenius theorem  for non-negative
matrices is used to  get a lower limit for the maximum eigenvalue of
the matrix $C_{nm}$:
\vspace{.3cm}

${\it Theorem}$: Suppose C is an $n\times n$
 non-negative primitive
matrix\footnote{A square non-negative matrix $T$ is said to be
primitive if there exist a positive integer $k$ such that $T^k>0$.}\footnote{
The extension from the finite to the countable case is possible \cite{se} and
necessary in our problem, but
will not be given explicitly  for reasons of simplicity.} .
Then there exists an eigenvalue $\lambda_{m}$ such that:

(a) $\lambda_{m}>0$;

(b)  $\lambda_{m}>\lambda$ for any eigenvalue  $\lambda_{m}\neq \lambda$;

(c) the eigenvectors associated with $\lambda_{m}$ are unique up to
constant multiples.

A proof can be found in \cite{se}.

The maximum eigenvalue can be calculated in the following way:
\begin{eqnarray}
\lambda_{max}=\sup_{\vert\vert x\vert \vert =1} \min_{i}
\frac{\sum_{j}c_{ij}x_{j}}{x_{i}}
\end{eqnarray}
This supremum is attained for some sequence $x$ with $l^2$-norm equal to
one and every element
 unequal to zero.

For example,  setting all $a_{n}$ equal produces

\begin{eqnarray}
\lambda_{max}\geq\frac{Si(A)}{\pi}
\end{eqnarray}
for $N\rightarrow \infty$.

As a result we now have an upper and lower bound for
the maximum probability in the physically most  relevant case.
Expanding our solutions also gives us   information about
 $p_{max}$:
\begin{eqnarray}
 p_{max}= \frac{A}{\pi}-O(A^3) \;\;\;\;\;\; \rm{for} \;\;A<\frac{\pi}{2}\, .
\end{eqnarray}
A further improvement of the bounds, omitted for reasons of brevity, can be
gained by using variational
methods.
Improvement of the bounds can also be gained by exploiting the
following theorem:

If $r$ is the Perron-Frobenius eigenvalue of an
irreducible\footnote{An $n\times n$ non-negative matrix $c_{ij}$ is
irreducible if for every pair $i,j$ of its index set, there exists a
positive integer $m\equiv m(i,j)$ such that $c^{(m)}_{ij}>0$}
 matrix
$\{c_{ij}\}$, then for any vector $x\in P$, where $P=\{x;x>0\}$:
\begin{eqnarray}
\min_i \frac{\sum_j c_{ij} x_j }{x_i}\leq r\leq \max_i\frac{\sum_j c_{ij} x_j
}{x_i}\nonumber
\end{eqnarray}
 This was proven by Collatz \cite{co} in 1942.


\vspace{1cm}
\subsection{Young's-Inequality}

The inequality
\begin{eqnarray}
\int dx\int dy
f(x)g(x-y)h(y)<D_{rst}\vert\vert f\vert\vert_r\,\,\vert\vert
g\vert\vert_s\,\,\vert\vert
h\vert \vert_{t}
\end{eqnarray}
proven independently  by  Beckner \cite{br} and by Brascamp-Lieb \cite{br}  is
based on Young's inequality
and gives  another way to calculate an
upper bound for $p(\sigma_x,\sigma_y$,T).

The relevant theorem is from the paper of Brascamp-Lieb.

\vspace{.1cm}

{\it Theorem:}  For $f\in L^{r}$, $g\in L^{s}$ ,  $h\in L^{t}$,
$r,s,t\geq 1$ and $\frac{1}{r}+\frac{1}{s}+\frac{1}{t}=2$, then
\begin{eqnarray} |\int dx\int dy
f(x)g(x-y)h(y)|<D_{rst}\,||f||_r\,\,||g||_s\,\,||h||_{t}
\end{eqnarray} with
$D_{rst}=r^{\frac{1}{r}}/r'^{\frac{1}{r'}}\,\,s^{\frac{1}{s}}/s'^{\frac{1}{s'}}\,\,t^{\frac{1}{t}}/t'^{\frac{1}{t'}}$
and  $r'=(1-\frac{1}{r})^{-1}$.

The inequality is only sharp if $f,g,h$ are certain types of Gaussian.
In our case $r,t=2$, $s=1$ and
\begin{eqnarray}
f(x)&=&\psi(x)\Theta (x+\frac{1}{2})\Theta (\frac{1}{2}-x)\nonumber\\
g(z)&=&\frac{\sin(Az)}{\pi z}\Theta (z+1)\Theta (1-z)\nonumber\\
h(y)&=&\psi^{*}(y)\Theta (y+\frac{1}{2})\Theta (\frac{1}{2}-y)\, .\nonumber
\end{eqnarray}
This leads to
\begin{eqnarray}
p(\sigma_x,\sigma_y,T)\leq \frac{2}{\pi}{\it Si}(A)
\end{eqnarray}
if $A<\pi$.

\vspace{1cm}

\section{Position and Momentum Samplings}
Now we use methods developed in section III to calculate the maximum
probability in the case of a
position-momentum history.
First we have to rewrite the probability-equation in an usable form:
\begin{eqnarray}
p(\sigma_k,\sigma_x,T)&=&
 \frac{1 }{ 2 \pi \hbar}
\int_{-\frac{\sigma_k}{2}}^{\frac{\sigma_k}{2}} dk
\int_{-\frac{\sigma_x}{2}}^{\frac{\sigma_x}{2}}
dx \int_{-\frac{\sigma_x}{2}}^{\frac{\sigma_x}{2}} dx'
 \langle x|e^{iHT}|k\rangle \langle k|e^{-iHT}|x'\rangle
  \psi(x)\psi^*(x')\nonumber\\
&=&\frac{A}{\pi}\int_{-\frac{\sigma_x}{2}}^{\frac{\sigma_x}{2}}dx\int_{-\frac{\sigma_x}{2}}^{\frac{\sigma_x}{2}}dx'\int_{-\frac{1}{2}}^{\frac{1}{2}}dk
\,\, e^{iA(x-x')k}
 \psi (x) \psi^* (x')\nonumber\\
&=&\frac{1}{\pi}\int_{0}^{\sigma_x}dx \int_{0}^{\sigma_x}dx'\,\,
\frac{\sin(A(x-x'))}{x-x'}
 \psi (x) \psi^* (x')
\end{eqnarray}
with $A=\sigma_k \sigma_x/2 \pi \hbar$.
This time no redefinition of the wave-function is necessary and we can
discretize $\psi$ directly.
The rest of the calculation is now identical to section II.
The result is an upper bound for the  probability:
\begin{eqnarray}
 p(\sigma_k, \sigma_x,T)\leq \frac{2 Si(\frac{A}{2})}{\pi}.
\end{eqnarray}

\subsection{Local Uncertainty Relation}
An  upper bound can also  be calculated
using the local uncertainty relation developed by Price and Faris.
First  a general inequality proven by Price \cite{pr}:

{\it Theorem:} Suppose $E\subseteq \Re^d $ is measurable and
$\alpha>\frac{d}{2}$. Then
\begin{eqnarray}
\int_{E}|\hat{ f}(k)|dk<K_{1} \, m(E)\, ||f||^{2-\frac{d}{\alpha}}_{2}
\end{eqnarray}
and
\begin{eqnarray}
\int_{E} dk|\hat{ {f}}|^2\leq  {\rm const}\,\, {\it m}(E)\,\, \vert\vert
f\vert\vert_{2}^{2-\frac{d}{\alpha}}\,\,\vert\vert\,|x|^{\alpha}\,f\vert\vert
_{2}^{\frac{d}{\alpha}}
\,\,\vert\vert \,|x|^{\alpha}\,f\vert\vert^{\frac{d}{\alpha}}_{2}
\end{eqnarray}
for all  $f\in L^{2}(R)$, where
\begin{eqnarray}
K_{1}=\frac{\theta(d)}{2 \alpha}\Gamma\Big(\frac{d}{2
\alpha}\Big)\Gamma\Big(1-\frac{d}{2 \alpha}\Big)\,\,\Big(\frac{d}{2
\alpha}-1\Big)^{\frac{d}{2\alpha}}\,\,\Big(1-\frac{d}{2
\alpha}\Big)^{-1}\, .
\end{eqnarray}
\vspace{.3cm}
In our special case we have
\begin{eqnarray}
\int_{E} |\hat{f}(k)|^2 dk \leq 2 \pi\,\, {\it m}(E)\,\, ||f||_2\,\,||xf||_2\,
\, .
\end{eqnarray}

We assume $f=\psi$ and
$||\,|x|\,\psi ||_{2}\leq \frac{\sigma_x}{2}||\psi ||_{2}$  to get
\begin{eqnarray}
p(\sigma_x,\sigma_k,T)\leq \frac{\sigma_x \sigma_k}{2 \pi \hbar }\, .
\end{eqnarray}

We do not have to study the momentum-position-case independently because of
time
reversibility in quantum mechanics.  In the inequality this just means that the
Fourier transform of
$\psi$ replaces the original wave-function.

\vspace{.2cm}
\section{Two-Momentum Histories}
Now we calculate the probability of a two-momentum history.
In the case of free-particle propagation the result is trivial,
because momentum is a conserved quantity.
The simplest case worth studying is the harmonic oscillator.
First we have to rewrite the probability-equation for the harmonic
oscillator in a simpler form:

\begin{eqnarray}
p(\sigma_{k'},\sigma_k,T)&=&
\int_{-\frac{\sigma_{k'}}{2}}^{\frac{\sigma_{k'}}{2}}dk'
\int_{-\frac{\sigma_k}{2}}^{\frac{\sigma_k}{2}}dl
\int_{-\frac{\sigma_k}{2}}^{\frac{\sigma_k}{2}}dl'
\hat{K}(l,T;k',0)
 \hat{K}^*(l',T;k',0)\hat{\psi}(l) \hat{\psi}^{*}(l')\nonumber\\
&=&\int_{-\frac{\sigma_{k'}}{2}}^{\frac{\sigma_{k'}}{2}}dk'
\int_{-\frac{\sigma_{k}}{2}}^{\frac{\sigma_k}{2}}dl
 \int_{-\frac{\sigma_k}{2}}^{\frac{\sigma_k}{2}}dl'
   \frac{- \sin (\omega T)}{\pi m \omega  \hbar }
\exp \Big\{
\frac{-i\sin(\omega T)   }{\hbar\omega
m}
[l^2-\frac{2k'l}{\cos (\omega T)}]\Big\}\nonumber\\
&&
\exp \Big \{
\frac{i\sin(\omega T)  }{\hbar\omega
m }
[l'^2-\frac{2k'l'}{\cos (\omega T)}]\Big\}
\hat{\psi}(l) \hat{\psi}^*(l')\nonumber
\end{eqnarray}

This time a simple  redefinition of the wave-function is necessary
before we can
use the methods developed previously directly.
The  upper bound for the probability is given by
\[p(\sigma_{k'}, \sigma_k,T)\leq\frac{2}{\pi} Si\Big(\frac{\sigma_{k}
\sigma_{k'}\sin(\omega T)
 }{2\omega m\hbar  }\Big)\]

\section{Two-projection histories for other Lagrangians}
In the next few subsections the methods developed above are applied to
other Lagrangians. In all the cases we analyse, the maximum probability remains
translation
invariant. This can be
seen directly  by just redefining our wave-function appropriately.

\subsection{Harmonic oscillator}
For the harmonic oscillator the Lagrangian has the form
\begin{eqnarray}
L=\frac{1}{2} m \dot{x}^2-\frac{1}{2} \omega^2 m^2 x^2
\end{eqnarray}
The corresponding Green's function (propagator) is
\begin{eqnarray}
K(x,T;y,0)=\sqrt{\frac{m \omega}{2 \pi i \hbar \sin(\omega
T)}}\exp\{\frac{im\omega }{2\hbar \sin(\omega
T)}[(x^2+y^2)\cos(\omega T)-2xy]\}
\end{eqnarray}
We will show that the   probability for a history in this potential is
\begin{eqnarray}
p(\sigma_x,\sigma_y,T)\leq\frac{2}{\pi} Si\Big(\frac{m \omega
\sigma_x \sigma_y}{4 \hbar \sin(\omega T)}\Big)
\end{eqnarray}

The derivation is sketched in the next few lines and is again based on
methods developed in the previous sections.
The probability is
\begin{eqnarray}
p(\sigma_x,\sigma_y,T)&=&\int_{-\frac{\sigma_x}{2}}^{\frac{\sigma_x}{2}}dx\Big
\vert
\int_{-\frac{\sigma_y}{2}}^{\frac{\sigma_y}{2}}dye^{iA(x^2+
y^2)\cos(\omega T)-2xy}\psi(y,0)\Big\vert^2
\nonumber\\
&=&
\int_{-\frac{\sigma_x}{2}}^{\frac{\sigma_x}{2}}dx
\int_{-\frac{\sigma_y}{2}}^{\frac{\sigma_y}{2}}dy
\int_{-\frac{\sigma_y}{2}}^{\frac{\sigma_y}{2}}dy'
e^{i2Ax(y'-y)}\psi_{new}(y,0)\psi_{new}^{*}(y',0) \, .
\end{eqnarray}
There are two  points of interest:
 To avoid having to evaluate complicated integrals,
it is again necessary to redefine  the
wave-function. Secondly, the similarity  between the maximal probability $\psi$
in
the free-particle case and the  case just described  allows us to
generalize our method to all linear cases.

\subsection{Constant Electric-field}
In this case of a constant electric-field the Lagrangian has the form:
\begin{eqnarray}
L=\frac{1}{2}m\dot{x}^2-ex\, .
\end{eqnarray}
And the relevant propagator is
\begin{eqnarray}
K(x,T;y,0)=\sqrt{\frac{m }{2 \pi i \hbar
T}}\exp\Big\{\frac{i}{\hbar}[\frac{m(x-y)^2}{2
T}-\frac{1}{2}e T(x+y) -\frac{e^2T^3}{24 m}]\Big\}\, .
\end{eqnarray}
The probability for history in this potential is accordingly
\begin{eqnarray}
p(\sigma_x,\sigma_y,T)\leq\frac{2}{\pi}Si \Big( \frac{m
\sigma_x \sigma_y}{4  \hbar  T}\Big)\, .
\end{eqnarray}
\subsection{Time-dependent Electric-field and Harmonic Oscillator}
As a generalization of this case we can also calculate
the maximum probability for the Lagrangian
\begin{eqnarray}
L=\frac{1}{2}m\dot{x}^2-e(t)x-\frac{m \omega^2}{2}x^2
\end{eqnarray}
where the  propagator is
\begin{eqnarray}
K(x,T;y,0)=\sqrt{\frac{m \omega}{2 \pi i \hbar \sin(\omega
T)}}
\exp\Big\{(\frac{i m \omega}{2 \hbar\sin(\omega T)})
[\cos(\omega
T)(x^2+y^2)
- 2 x y\nonumber\\- 2\frac{y}{m \omega} \int_{0}^{T}dt e(t) \sin(\omega
(t-T)) -2\frac{x}{m \omega} \int_{0}^{T}dt e(t)\
\sin(\omega
(T-t))\nonumber\\ -2\frac{1}{m^2 \omega^2}\int_{0}^{T}dt \int_{0}^{T}dt'
e(t)e(t')
\sin(\omega
(t-T)\sin(\omega(t'-T))]\Big\}\, .\nonumber
\end{eqnarray}
The calculation is identical to the previous sub-section, except that we
have to redefine $\psi $ to $\psi e^{iAy^2}e^{2\frac{y}{m \omega}\int_{0}^{T}dt
e(t) \sin(\omega
(t-T))}$.

The resulting  probability is
\begin{eqnarray}
p(\sigma_x,\sigma_y,T)\leq \frac{2}{\pi}Si\Big(\frac{m \sigma_x
\sigma_y \omega}{4 \hbar
sin(\omega t)}\Big)
\end{eqnarray}
for $m \omega\sigma_x\sigma_y/\hbar
sin(\omega t)>\frac{\pi}{2}$.
As in all the previous cases  a lower bound for the maximal
probability can be computed which is
\begin{eqnarray}
p_{max}(\sigma_x,\sigma_y,T)\geq \frac{1}{\pi} Si\Big(\frac{m
\sigma_x
\sigma_y
\omega}{2 \hbar
sin(\omega t)}\Big)\, .
\end{eqnarray}
This means that  any influence by an electrical field can be counteracted by
choosing  appropriate initial conditions.
\subsection{General Lagragians}
In the case of arbitrary linear systems
the propagator has the form
\begin{eqnarray}
\langle x'',t''\vert x',t'\rangle = \Delta
(t'',t')\exp\Big\{\frac{i}{\hbar}S(x'',t''\vert x',t')\Big\}\, .
\end{eqnarray}
A corresponding upper bound for the probability is then
\begin{eqnarray}
p(\sigma_x,\sigma_y,T)\leq \sigma_x \sigma_y |\Delta (t'',t')|^2
\end{eqnarray}
which follows directly from the H\"older inequality\footnote{If
$f\in L^p $ and $g \in L^{p'} $ with $\frac{1}{p}+\frac{1}{p'} =1$ then: $\int
dx \,\,
f(x)\,g(x)\leq \vert \vert f \vert \vert_p\,\,  \vert \vert g \vert
\vert_{p'}$}.
For reasons of brevity no further examples are given.
It should be clear by now that a very large class of physical
problems can be studied by the methods developed in the
previous sections.

\section{N-projections}
We can use the same generalizations to give
a simple improvement  to the N-projection probability given in \cite{jjh1}.
At first we assume that the N-projection history is made up out of $N-1$
two-projection histories.
The new upper bound for the probability is then:
\begin{eqnarray}
p(\sigma_{x,1},...,\sigma_{x,n})\leq
\Big(\frac{2}{\pi}\Big)^{N-1}\prod_{n=1}^{N-1}
Si\Big(\frac{\sigma_{x,n} \sigma_{x,n+1} m}{4 \pi \hbar T_{ij}}\Big)
\end{eqnarray}
where  $T_{ij}$ is the time difference between the successive projections.
Analogous  upper bounds for the probability can be given for all the
other cases studied before.

\section{Information Theory}


An lower bound for the information is given by
\begin{eqnarray}
I\geq - \log(p_{max})
\end{eqnarray}
because of $\sum p=1$.

In our case this gives an lower bound for the information of
\begin{eqnarray}
I\geq -\log \Big(\frac{2Si( \frac{A}{2})}{\pi}\Big)\, .
\end{eqnarray}
This is true in all the examples described above.
The constant $A$ is dependent on the physical situation and has been
given in previous sections.

Next we show how one can calculate an  $I_{d}$ for different conventionally
used
wave-functions.
At first
we look at two-position histories.
We assume $\psi$ to be  divided again in piecewise constant
functions with
\[
\psi_{n}(x)=\left\{\begin{array}{ll}
\sqrt{\sigma_y}&\mbox{$n\sigma_y\leq x\leq (n+1)\sigma_y$}\\
0&\mbox{otherwise}
\end{array}
\right.
\]
and
\begin{eqnarray}
\sum_n |a_n|^2=1\, .
\end{eqnarray}
The transition probability is
\begin{eqnarray}
p(n,m)=\frac{A}{\pi}\int_{\sigma_x
n}^{\sigma_x(n+1)}dx\int_{\sigma_ym}^{\sigma_y(m+1)}dy
\int_{\sigma_ym}^{\sigma_y(m+1)}dy'
\exp\Big\{i\frac{2Ax(y-y')}{\sigma_x\sigma_y}\Big\}
\psi(y)\psi^{*}(y')\, .
\end{eqnarray}
The fact   that $\psi_{n}$ is constant in each m-interval,
leads to
\begin{eqnarray}
p(n,m)=p'(n)p(m)\,\, ,
\end{eqnarray}
\begin{eqnarray}
p(m)=|a_{m}|^2
\end{eqnarray}
and
\begin{eqnarray}
p'(n)&=&\frac{A}{\pi}\int_{n}^{n+1
}dx\int_{0}^{1}dy\int_{0}^{1}dy'e^{i2Ax(y-y')}\nonumber\\
&=&\frac{A}{\pi}\int_{0}^{1}dx\int_{0}^{1}dy\int_{0}^{1}dy'e^{i2A(x+n)(y-y')}\nonumber\\
&=&\frac{1-(n+1)\cos(2An)+ n\cos(2A(n+1))}{2\pi
n(n+1)A}+\frac{Si(2A(n+1))-Si(2nA)}{\pi}\nonumber
\end{eqnarray}
for $n>0$.
The corresponding information is
\begin{eqnarray}
I_{d}&=&-\sum p(n,m)\log p(n,m)\nonumber\\
&=&-\sum p(n)p'(m)\log(p(n)p'(m))\nonumber\\
&=&-\sum_{n} p(n)\log p(n)+p'(n)\log p'(n)\,
\end{eqnarray}


For any  $\sigma_y$ and $\sigma_x$ and given $a_{n}$ we can now
calculate the information explicitly.
This sum has its minimum when $p(n)=\delta_{nm}$.

Another interesting case involves setting $\sigma_y$ and $\sigma_x$ to
be much less than one, thereby approximating the Shannon information  required
for the continuous case.
The information is
\begin{eqnarray}
I_{min,d}=-\sum_{n} p'(n)\log p'(n)\, .
\end{eqnarray}

For other two-projection histories the process is similar.

\section{Discussion and Outlook}

In this paper several mathematical methods are developed to
calculate the maximum probabality and minimum Shannon information of histories
under varying
conditions. Not only   upper but also  lower bounds for the maximum
probability are calculated. Special importance is placed on the
 similarities between the histories for different Lagrangians.
Next   a lower  bound for the
corresponding information is calculated.

These results can be used to do explicit calculations in the
framework of decoherent histories and to construct experiments to verify
quantum
mechanics, by comparing the maximal  experimentally achievable probability
with our
theoretical  bounds.

This paper also can be used to support the view that Gaussian slits
are not as unphysical as normally assumed, but can have experimental
relevance. This can be shown by comparing the maximal probabilities
and lower bounds for the Shannon information
for Gaussian, as described in \cite{jjh1}, and exact projections in a variety
of situations.



There are three major areas where the formalism should be extended:

\begin{itemize}
\item There is the need to gain an exact bound for the maximal
probability and Shannon information
for a general N-projection history.

\item One needs to generalize  the results from one to three
spacial  dimensions.

\item One needs to study projection smeared in time.
\end{itemize}
This will be done in a forthcoming paper.

\section{Acknowledgement}

I want to thank J. J. Halliwell and C. J. Isham for many helpful suggestions
and comments
and  H.G.Feichtinger for his assistance during my stay at
the math-department of Wien University.
This work was supported  by  CVCP.


\begin{references}



\bibitem{ar} A. Anderson and  J.J. Halliwell, Phys. Rev. D \bf{48},
\rm 1580 (1992).
\bibitem{be} W. Beckner, Ann. of Math., \bf{102}\rm , 159-82 (1975).
\bibitem{br} H. J. Brascamp and E. H. Lieb, Advances in Math.
\bf{20} \rm , 151-173 (1976).
\bibitem{co} L. Collatz, Mathmathische Zeitschrift, \bf{48}\rm , 221-6  (1942).
\bibitem{co1} T. M. Cover and J. A. Thomas, \it{Elements of
Information Theory}\rm , (Wiley, New York, 1991).
\bibitem{da} I. Daubechies, \it{Ten Lectures on Wavelets}\rm ,  (SIAM,
Philadelphia, PA, 1992).
\bibitem{fa} W. G. Faris, J.
Math. Phy. $\bf{19}$,  461-466 (1978).

\bibitem{gr} U. Grenader and G. Szeg\"o,\it{T\"opltiz Forms and their
applications}\rm, \, (Univ. California Press, Berkley and L.A., 1958).
\bibitem{gri} R. Griffiths, J. Stat. Phy. \bf{36}\rm, 219, (1984);
M. Gell-Mann and J.B. Hartle, in \it{Complexity, Entropy and the
Physics of Information}\rm, edited by W. Zurek, SFI Studies in the
Science of Complexity Vol. VIII (Addison-Wesley, Reading, MA, 1990);
R. Omnes, Rev. Mod. Phys. \bf{64}\rm, 339 (1992).
\bibitem{jjh1} J.J. Halliwell, Phys. Rev. D \bf{48}\rm , 2739 (1993).
\bibitem{ha} J. B. Hartle, \it{Spacetime quantum mechanics and the
quantum mechanics of spacetime},\rm (Proceedings on the 1992 Les
Houches School, Gravitation and Quantisation,  1993).
\bibitem{is} C. J. Isham, IMPERIAL/TP/92-93/39, appearing in  J.
Math. Phy..
\bibitem{ka} M. Kac, W.L. Murdock \& G. Szeg\"o, Journal of Rational
Mechanics and Analysis, Vol. 2, No. 4, Oct. 1953,  767-799.

\bibitem{me} Y. Meyer, \it{Wavelets and Oerators}\rm, (Cambridge University
Press , 1992).

\bibitem{pa} M. H. Partovi, Phys.Rev.Lett. ${\bf 50}$, 1883(1983).
\bibitem{pr} J. F. Price,
Studia Mathematica \bf{85}\rm , \, 26-45 (1987).
\bibitem{sc} I. Schur, \"Uber einen Satz von Caratheodory,
Sitzungsbericht der Kgl. Preussischen Akademie der Wissenschaften,
1912,  4-15.
\bibitem{se} E. Seneta, \it{Non-Negative Matrices}\rm, \, (George Allen \&
Unwin, London, 1973).
\bibitem{sh} C.E. Shannon and W.W. Weaver, The Mathematical Theory
of Communication, (University of Illinois Press, Urbana, IL., 1949).
\bibitem{sz} G.Seg\"o, Mathematische Zeitschrift \bf{6}\rm, \,
167-202 (1921).
 \end{references}
\end{document}